\documentclass[prb,twocolumn,showpacs]{revtex4}
\usepackage{epsfig}
\usepackage{graphicx}

\newcommand{\be} {\begin{equation}}
\newcommand{\ee} {\end{equation}}
\newcommand{\bea} {\begin{eqnarray}}
\newcommand{\eea} {\end{eqnarray}}

\begin{document}
\preprint{HEP/123-qed}
\title{Mean-Field Theory, Mode-Coupling Theory, and the Onset Temperature in Supercooled Liquids}
\author{Yisroel Brumer and David R. Reichman}
\affiliation{Harvard University, 12 Oxford Street, Cambridge, Massachusetts 02138}
\date{\today}
\begin{abstract}
We consider the relationship between the temperature at which averaged
energy landscape properties change sharply ($T_{o}$), and the
breakdown of mean-field treatments of the dynamics of supercooled
liquids.  First, we show that the solution of the wavevector dependent
mode-coupling equations undergoes an ergodic-nonergodic transition
consistently close to $T_{o}$.  Generalizing the landscape concept to
include hard-sphere systems, we show that the property of inherent
structures that changes near $T_{o}$ is governed more fundamentally by
packing and free volume than potential energy.  Lastly, we study the
finite-size Random Orthogonal Model (ROM), and show that the onset of
noticeable corrections to mean-field behavior occurs at $T_{o}$.
These results highlight new connections between the energy landscape
and mode-coupling approach to supercooled liquids, and identify what
features of the relaxation of supercooled liquids are properly
captured by mode-coupling theory.
\end{abstract}
\maketitle

The slow dynamics that supercooled liquids undergo as they approach
the glass transition has defied a satisfying explanation for many
decades \cite{Angell, StillDeben}.  Several theoretical paradigms have
been presented that shed light on certain features of these dynamics.
The notion of an energy landscape has been useful for understanding
thermodynamic properties of the glassy state as well as for
rationalizing the connection between transport properties and the
packing structures associated with local potential energy minima
(``inherent structures'') that are visited at a given temperature
\cite{StillDeben, SastryNature}.  Other theories, such as
mode-coupling theory (MCT) \cite{Bengtzelius, Leut2},
have also been influential in explaining the sequence of relaxation
events that occurs in mildly supercooled liquids.
%For example, MCT has recently
%predicted the existence of logarithmic $\beta$-relaxation in colloidal
%systems with attractive interactions \cite{Gotze, Gotze2, Bergenholtz, Gotze3,
%Gotze4, Mallamace, Fabbian, Fabbian2, Fuchs, Fuchs2}, which has been
%verified by light scattering experiments \cite{Pham, Mallamace2, Weeks, Krall}
While successful in several contexts, both the landscape and MCT
approaches suffer from problems that limit their utility.  The
landscape picture does not provide a predictive, quantitative
framework for describing the dynamics of supercooled liquids.
Furthermore, it is not clear how one could apply energy landscape
concepts in a useful way to entropically dominated glassy systems such
as hard-sphere liquids.  MCT does provide such a framework, but
several important predictions made by this theory, including the
thermodynamic location of the ergodic-nonergodic transition, are
incorrect \cite{Nauroth}.  Here we elucidate connections
between these two viewpoints that shed new light on various features
of these seemingly different approaches.

Effort has been made to connect the intuitively based landscape
picture to the more mathematical mode-coupling approach.  Pioneering
work of Kirkpatrick, Wolynes and Thirumalai \cite{Kirk,KirkThir}
showed that the mode-coupling equations are exact for certain
mean-field spin glasses.  In particular, the $p=3$ $p$-spin model
exhibits a dynamical transition at a temperature $T_{c}$ and a
thermodynamic transition at a lower temperature $T_{K}$ \cite{Kirk2,
Sommers1, Sommers2, Sommers3}.  The temperature $T_{c}$ is associated
with a mean-field divergence of barriers, leading to nonergodicity.
It has been argued that effects beyond the mean-field limit render the
barriers at $T_{c}$ finite \cite{Kirk2, CrisantiRitort}.
Thus, $T_{c}$ corresponds to the temperature at which activated
processes over finite sized barriers dominate transport.  Since MCT in
its simplest form neglects these activated processes, the full
wavevector dependent MCT equations exhibit a ``glass'' transition at
$T_{c} > T_{g}$ where $T_{g}$ is the calorimetric glass transition
observed in the laboratory.  Several computer studies have attempted
to strengthen the connection between activated processes on the energy
landscape and the temperature $T_{c}$
\cite{SastryNature,Cavagna,Broderix, Angelani}.  Recently, however,
simulations have shown that activated processes strongly influence
transport in supercooled liquids at temperatures significantly in
excess of $T_{c}$ \cite{Doliwa1,Doliwa2,Denny}.  Thus, the
physical meaning and relevance of $T_{c}$ for finite dimensional,
non-mean-field glassy systems remains unclear.

Several years ago, Sastry {\em et al.} \cite{SastryNature}
pointed out the existence of another characteristic landscape
temperature, the ``onset'' (or ``landscape influenced'') temperature
$T_{o}$.  $T_{o}$ may be significantly larger than $T_{c}$ as
calculated by power-law fits of diffusive data, and coincides with the
onset of nonexponential and non-Arrhenius relaxation in supercooled
liquids.  Sastry {\em et al.} \cite{SastryNature} found that
$T_{o}$ also marked the temperature at which averaged energy landscape
properties (such as the average value of the inherent structure
energy) show a sharp change as a function of temperature.  While much
work has focused on understanding the qualitative changes in dynamics
near $T_{c}$, very little work has been devoted to understanding the
meaning of $T_{o}$.  Here we demonstrate connections
between MCT, (and mean-field-like approaches in general) and the
changes in landscape properties that occur as the system is cooled
below $T_{o}$.

It has long been known that the location of the temperature at which
the full wavevector dependent MCT equations predict a loss of
ergodicity is significantly higher than $T_{c}$ as obtained by
power-law fits of the temperature dependence of transport
coefficients.  Here we show that the ergodic-nonergodic transition
actually occurs close to $T_{o}$.  We solve the wavevector dependent
MCT equations for the $2 \times 2$ matrix $ {\bf \ddot{F}}({\bf q},t)
$ with matrix elements $ F_{ij}({\bf q},t), $ where  \cite{Nauroth} \be
{\bf \ddot{F}}({\bf q},t) + {\bf\Omega}^2({\bf q}) {\bf F}({\bf q},t)
+ \int_0^t d\tau {\bf M} ({\bf q}, t - \tau){\bf \dot{F}}({\bf q},
\tau) = 0. \ee The frequency matrix $ {\bf \Omega}^2({\bf q}) $
is defined as \be [{\bf \Omega}^2({\bf q})]_{ij} = \frac{{\bf
q}^2k_BTx_i}{m_i} \sum_k \delta_{ik} [{\bf S}^{-1}({\bf q})]_{kj} \ee
and the mode-coupling approximation for the memory function is given
by \bea M_{ij}({\bf q}, t) = \frac{k_BT}{2nm_ix_j} \int \frac{d{\bf
k}}{(2\pi)^3} \sum_{\alpha \beta} \sum_{\alpha' \beta'}V_{i \alpha
\beta} ({\bf q, k}) \nonumber \\ V_{j \alpha' \beta' }({\bf q, q - k})
F_{\alpha \alpha'} ( {\bf k}, t) F_{\beta \beta'} ({\bf q - k}, t)
\eea where $ n $ is the particle density, $ x_i $ and $m_i $ are,
respectively, the concentration and mass of particle type $ i $ and $
{\bf S}({\bf q}) $ is the $2 \times 2$ matrix of partial structure
factors.  Expressions for the vertices $ V_{i \alpha \beta} $ can be
found in the literature \cite{Nauroth}.
\begin{figure}[ht]
\includegraphics[width=0.5\linewidth,height=4.6cm]{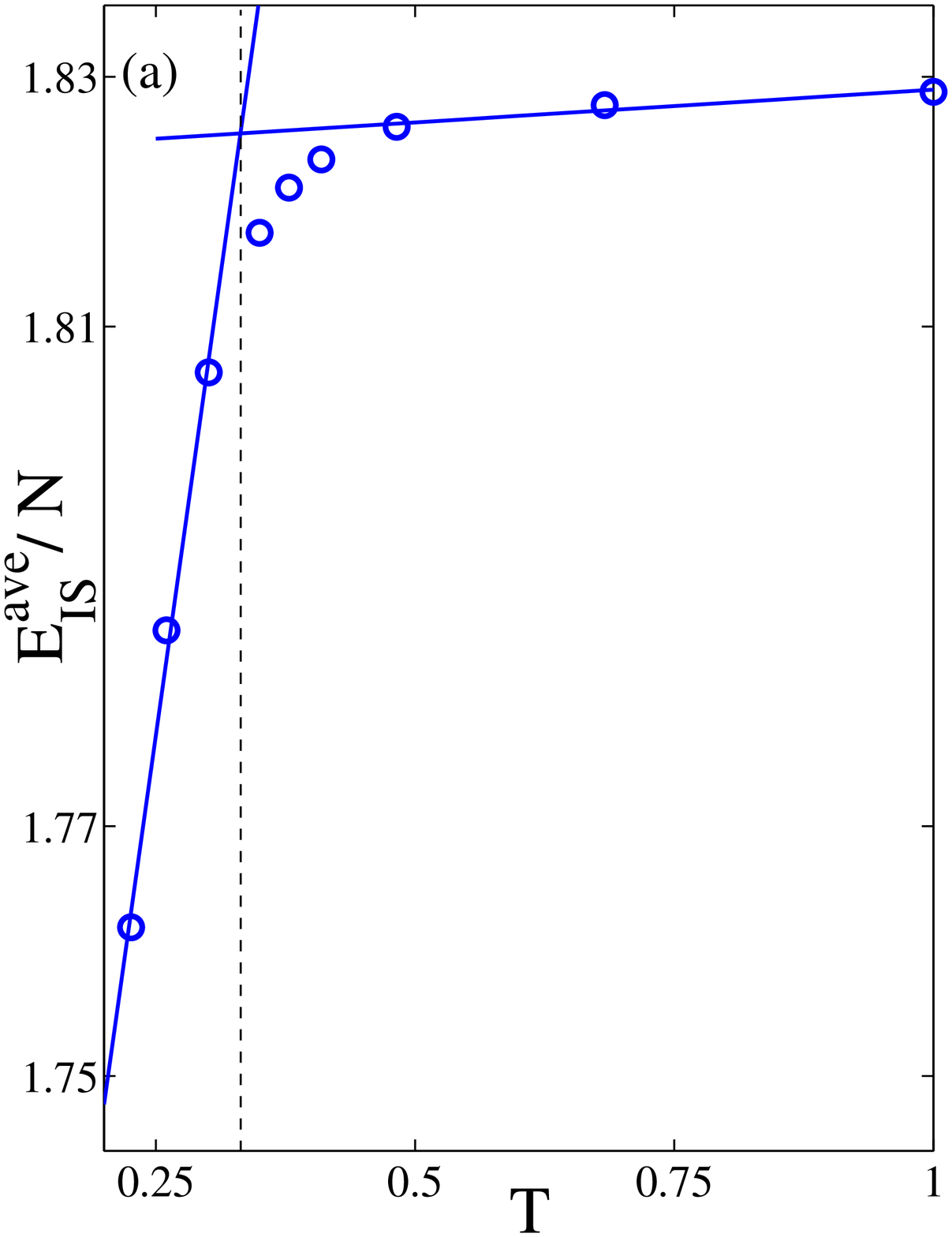}\hfill
\includegraphics[width=0.5\linewidth,height=4.6cm]{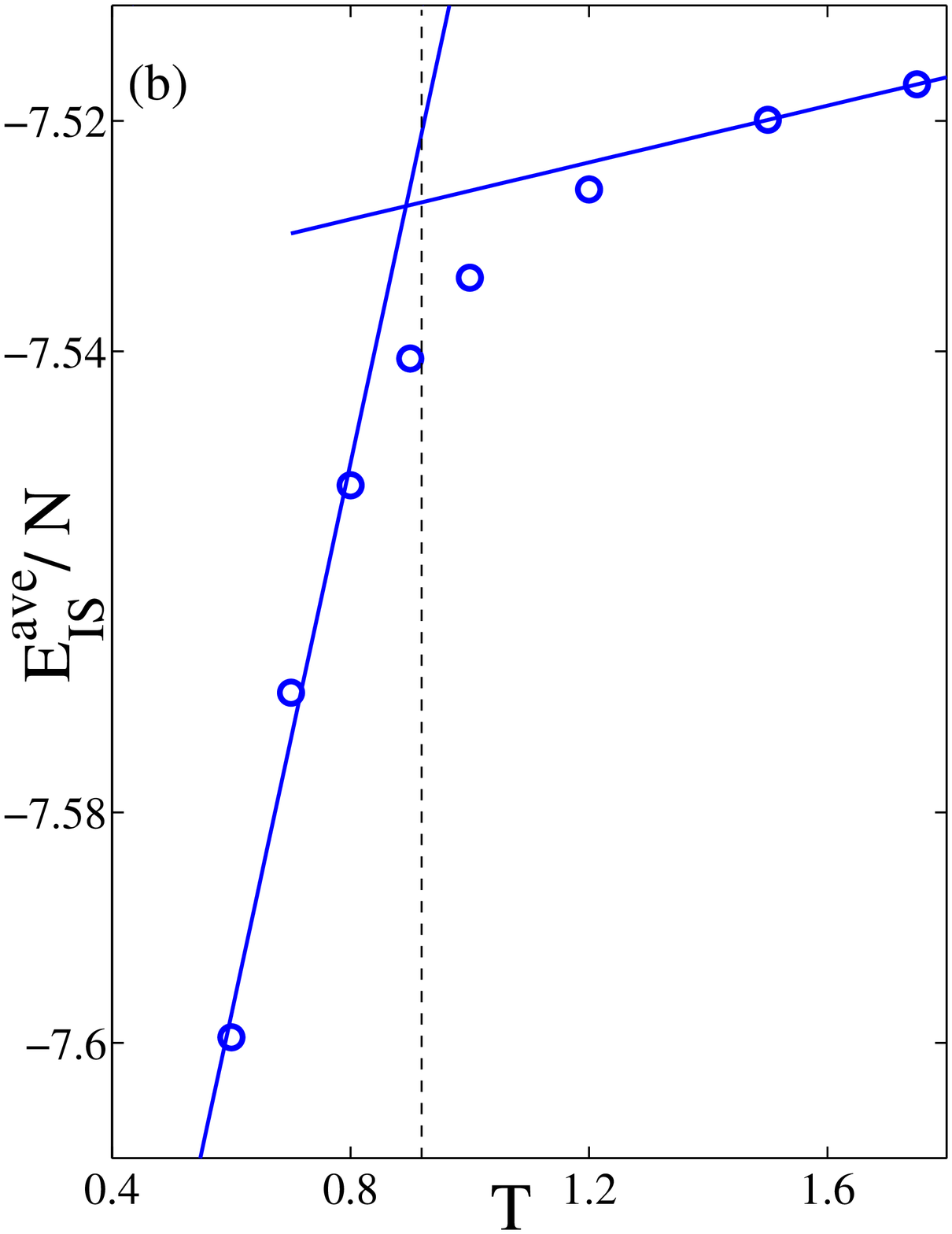}
\caption{Average inherent structure energy vs. temperature for (a) 70
particle 50/50 soft-sphere system and (b) 120 particle 80/20 LJ
system.  The solid lines are straight line fits to high and low
temperature slopes.  $ T_o $ is defined to be the temperature where
these lines cross. The vertical dashed line indicates the temperature
of the ergodic-nonergodic transition as calculated by Eqs. (1)-(3).
$ T_{c} $, as calculated by fits to diffusive data, is found to be $
T_c = 0.246 $ for the soft-sphere system and $ T_c = 0.435 $ for the
LJ system. Lennard-Jones units are used here and throughout the text.
Both systems are simulated at unit density.}
\end{figure}

Solving Eqs. (1)-(3) yields the location of the temperature at which
the function $F({\bf k},t)/S(k)$ fails to decay to zero as $t
\rightarrow \infty$.  In Fig. 1 we plot the average inherent structure
energy ($\overline{E}_{IS}(T)$) as a function of temperature for two
different potentials.  The first system studied is the 50/50
soft-sphere mixture studied by Barrat {\it et. al.} \cite{Barrat2}.
The second system is the 80/20 Lennard-Jones mixture studied by Kob
and Andersen \cite{KobAndersen}. The onset temperature $ T_o $ is
located where $\overline{E}_{IS}(T)$ show a sharp decrease as a
function of temperature, or where the 1st order polynomial fits to the
high and low temperature data cross.  A dashed line indicates the
location of the ergodic-nonergodic transition temperature obtained
from Eqs. (1)-(3).  Clearly, the location of the ergodic-nonergodic
transition occurs very close to the landscape onset temperature
$T_{o}$ and not $ T_c $ as calculated by fits to diffusion data.
Since MCT may be viewed as a particular type of dynamical mean-field
theory, the coincidence of the breakdown of MCT at $T_{o}$ signals the
failure of MCT to capture specific non-mean-field effects.  The nature
of this failure is discussed below.

A particularly troubling feature of energy landscape theories is the
inability to treat hard-sphere systems, where barrier crossing events
are completely entropic in nature.  In order to address this issue, we
generalize the inherent structure concept in a manner similar to that
discussed by Santen and Krauth\cite{Krauth}.  Specifically, we study a
binary hard-sphere system consisting of 225 particles with $ \sigma_1
= 0.1 $ and and 25 particles with $ \sigma_2 = 0.5 $ at various
packing fractions $ \eta$ \cite{Bosse2}. Starting with well
equilibrated configurations generated at a given packing fraction, the
system is ``crunched'' until hard-sphere overlap occurs.  Some Monte
Carlo moves are made during this process to ensure the system reaches
a stable packing structure.  The final, stable configuration obtained
from this procedure is called an inherent structure.  In Fig. 2a we
show the average inherent structure volume ($\overline{V}_{IS}(\eta)$)
versus packing fraction for the binary hard-sphere system.
Remarkably, an onset packing fraction $\eta_{o}$ may be defined from
the inherent structures that again coincides with the
ergodic-nonergodic critical packing fraction as found from
Eqs.(1)-(3).  This not only demonstrates the robustness of the
correlation between the dynamics of the ergodic-nonergodic transition
as found directly from MCT via Eqs. (1)-(3) and the onset temperature
(packing fraction) $T_{o}$ ($\eta_{o}$), but it also hints at a deep
connection between inherent structures labeled by potential energy,
and configurations defined by packing and free volume.  To strengthen
this connection, we reconsider a thermal system, namely the
soft-sphere system studied in Fig. 1.  Using a crunching procedure
(with an upper energy cutoff) similar to that used in the hard-sphere
system, we calculate $\overline{V}_{IS}(T)$ versus $T$.  In Fig. 2b we
show that the onset temperature for changes in $\overline{V}_{IS}(T)$
quantitatively coincides with $T_{o}$ as extracted from inherent
structure energies ($\overline{V}_{IS}(T)$) for the same system.  This
demonstrates that the changes in the inherent structures that are
sampled in the liquid range where dynamics become nonexponential are
associated with sharp changes in free volume and structural packing
motifs.  In fact, such quantities are more fundamental than the
quenched potential energy and allow for the generalization of
landscape-like concepts to athermal systems such as hard-spheres.
\begin{figure}[ht]
\includegraphics[width=0.49\linewidth,height=4.6cm]{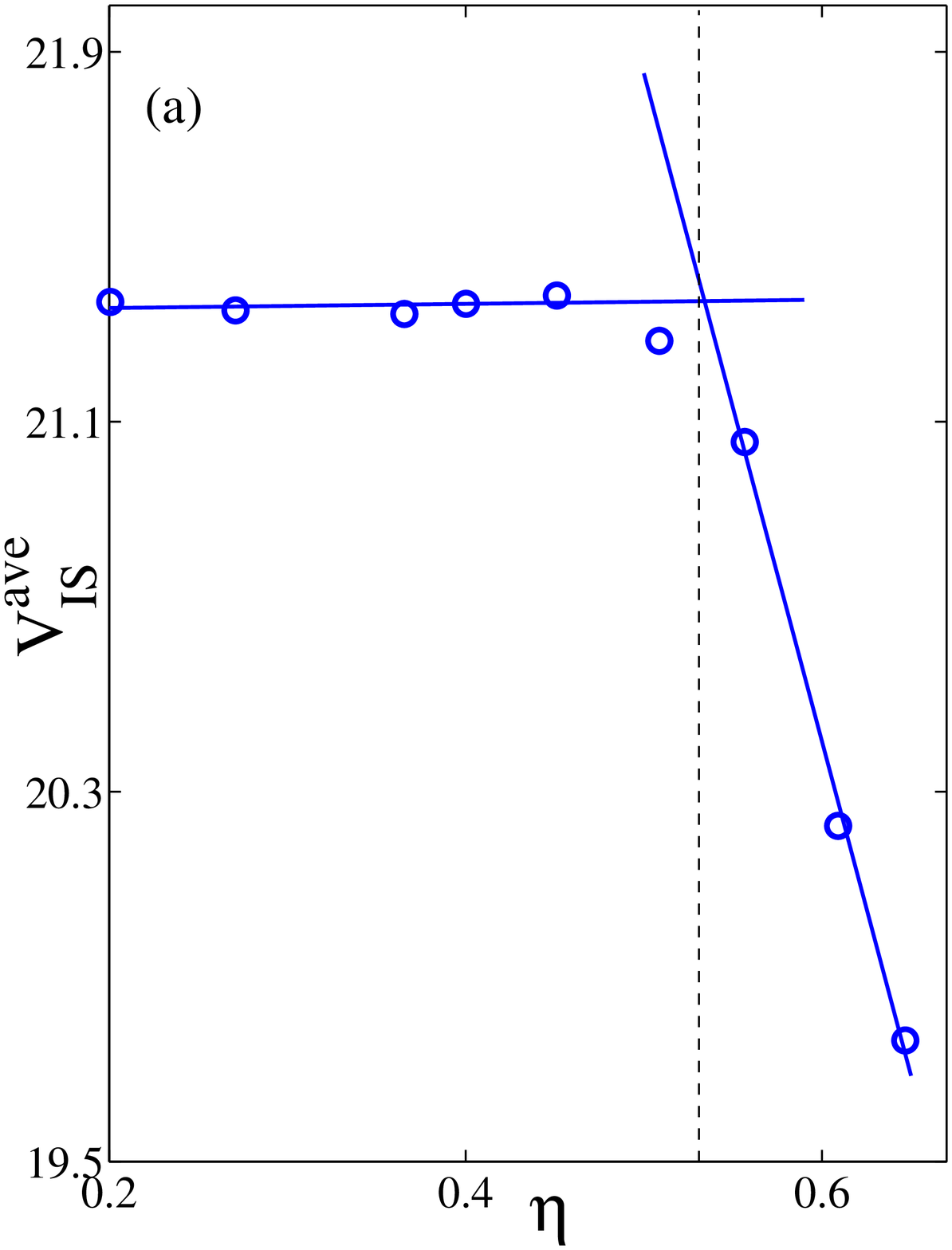}\hfill
\includegraphics[width=0.5\linewidth,height=4.6cm]{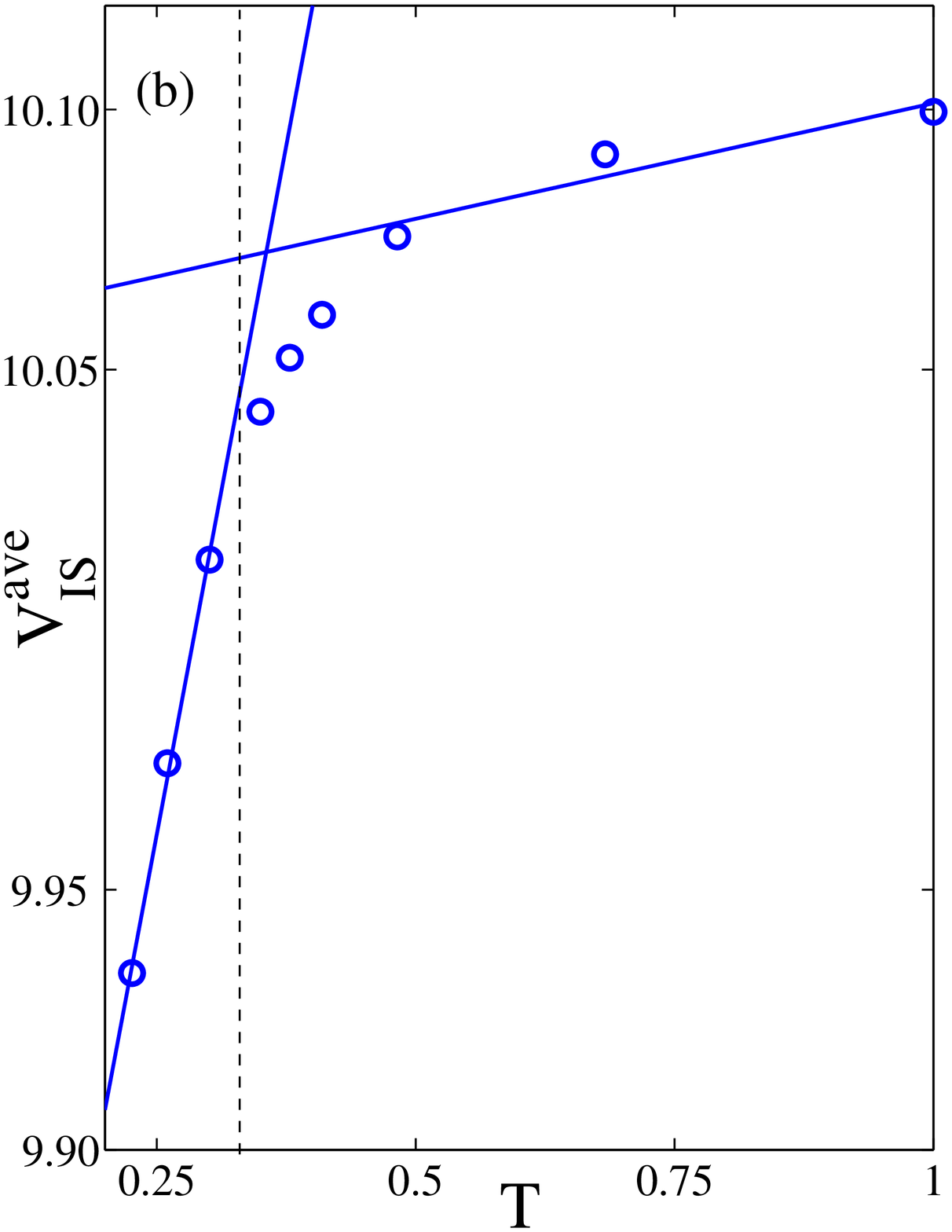}
\caption{(a) Average ``inherent structure'' volume vs. packing
fraction for the hard-sphere mixture as calculated by the
``crunching'' procedure described in the text.  The dashed line
indicates the location of the ergodic-nonergodic transition as
calculated from Eqs.(1)-(3). (b) ``Crunching'' procedure applied to the
soft-sphere system. An energy cutoff has been employed to define the
final IS volume.  The dashed line indicates the temperature of the
ergodic-nonergodic transition as calculated by Eqs.(1)-(3). $ T_o $ in
this system as defined via $ \overline{E}_{IS}(T) $, $ T_o^E = 0.332 $
(see Fig. 1a), agrees well with that extracted from the $
\overline{V}_{IS}(T) $ definition, $ T_o^V = 0.356$.}
\end{figure}

Despite the fact that the solution to the full set of Eqs.(1)-(3)
yields an ergodic-nonergodic transition temperature that is very close
to $ T_o $, one still expects MCT to give physically sensible results
in the range $ T_c < T < T_o $. Indeed, Kob {\em et al.} have demonstrated that
Eqs.(1)-(3) accurately account for dynamics at these temperatures if
the static input is calculated at a higher (effective) temperature\cite{Kob3},
and many of the scaling predictions of MCT appear to be corroborated
via computer simulations in this range \cite{Gotze,KobAndersen}.

The freezing of $F({\bf k},t)/S(k)$ at $T_{o}$ (and not $T_{c}$)
merely signals the {\em sensitivity} of MCT to changes in the packing
structure of the liquid and hints at the breakdown of the mean-field
like approximation that is inherent in the idealized version of
mode-coupling theory.  To get a better feeling for the nature of this
breakdown, we consider the dynamics of the random orthogonal model
(ROM) \cite{Marinari3}.  Specifically,
we take the Hamiltonian
\begin{equation}
H=-2 \sum_{ij} J_{ij} \sigma_{i} \sigma_{j},
\end{equation}
where $\sigma_{i}= \pm 1$ are Ising spin variables, and $J_{ij}$ is an
$N \times N$ random symmetric orthogonal matrix with $J_{ii}=0$. The
Glauber algorithm \cite{Glauber} is employed for dynamics. For $N
\rightarrow \infty$, this model is in the same dynamical universality
class as the $p=3$ spherical $p$-spin model.  Thus, in this limit, MCT
is exact.  For $N$ finite, $1/N$ corrections to the MCT should appear.
It has already been shown that for finite $N$, the model defined by
Eq.(4) behaves similarly to real liquids, exhibiting landscape
properties very similar to atomistic models, as well as non-trivial
dynamical properties such as Nagel scaling \cite{Rao}.  In
Fig. 3a a plot of $\overline{E}_{IS}(T)$ versus $T$ is shown for a
particular value of $N$.  The generic features of this plot are not
sensitive to $N$ over a wide range of values for finite $ N $.  An
onset temperature $T_{o} \approx 0.72$ is found for this model, while
$T_{c}=0.536$ and $T_{K}=0.26$. In Fig. 3b, we plot
$\chi_{N,N^{'}}(T)=\int_{0}^{\infty}dt|C_{N}(t)-C_{N^{'}}(t)|,$ where
$C_{N}(t)=\frac{1}{N}\sum_{i=1}^{N}\overline{\langle
\sigma_{i}(t)\sigma_{i}(0) \rangle}$ is the disorder-averaged
spin-spin correlation function for a finite $N$ system.  Clearly, this
``succeptability'' measures finite $N$ corrections to mean-field
dynamics for $N^{'}, N < \infty$.  In Fig. 3b, we plot
$\chi_{N,N^{'}}(T)$ versus $T$ for several different choices of $N$
and $N^{'}$.  Remarkably, $\chi(T)$ sharply increases from zero very
close to $T_{o}$.  This result is not sensitive to $N$ and $N^{'}$ for
a wide range of values.  Indeed, this result demonstrates that
noticeable corrections to mean-field (MCT) dynamics occur at $T_{o} >
T_{c}$ as calculated by fits to transport coefficients.
%Interestingly, we find no dramatic change in behavior in the plot of
%$\chi_{N,N^{'}}(T)$ versus $T$ near $T_{c}$ as long as $ N $ and $
%N^{'} $ are finite, implying that such effects vary {\em continuously}
%near the mode-coupling transition temperature for finite but arbitrary
%$ N^{'} > N $.
\begin{figure}[ht]
\includegraphics[width=0.5\linewidth,height=4.6cm]{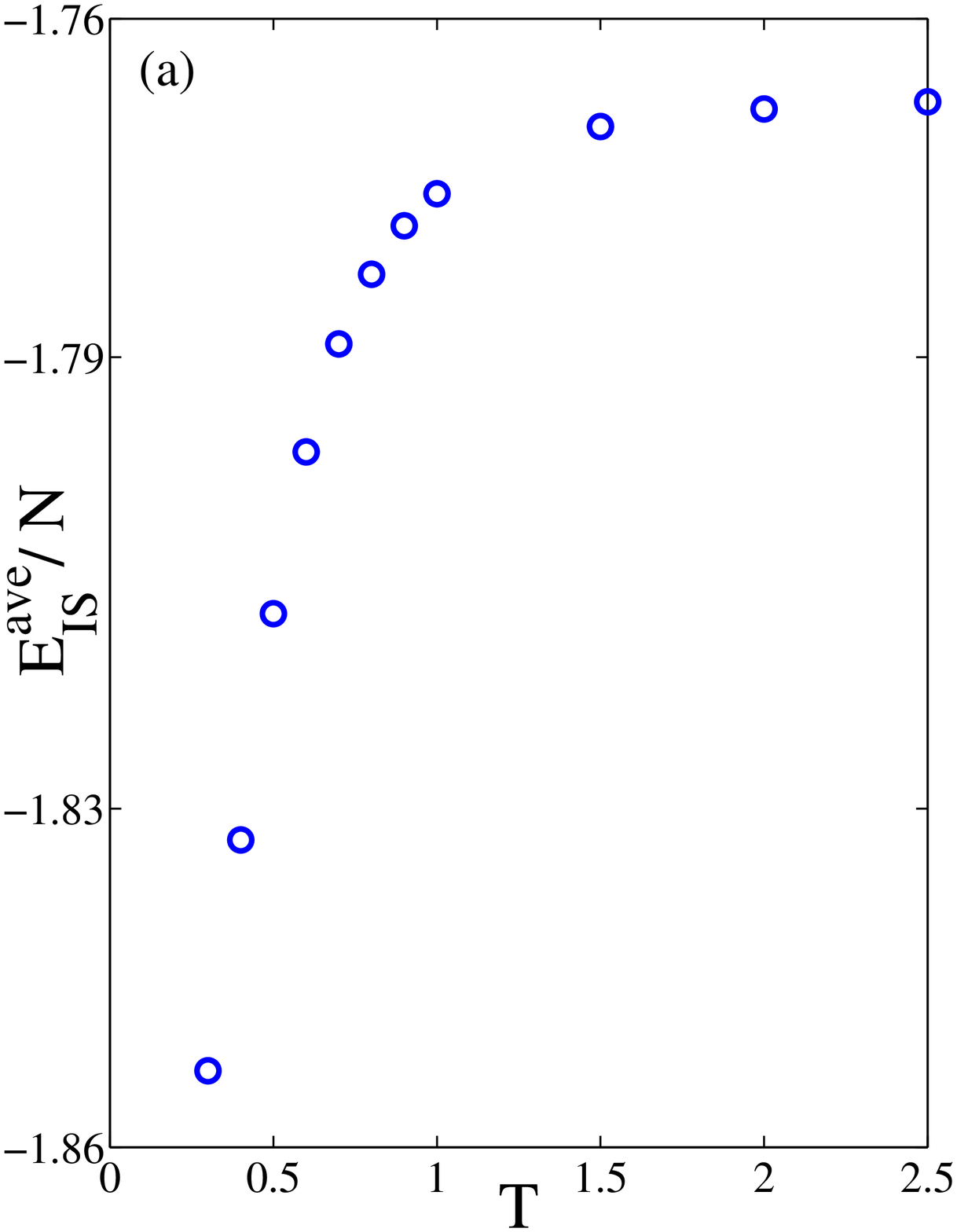}\hfill
\includegraphics[width=0.49\linewidth,height=4.6cm]{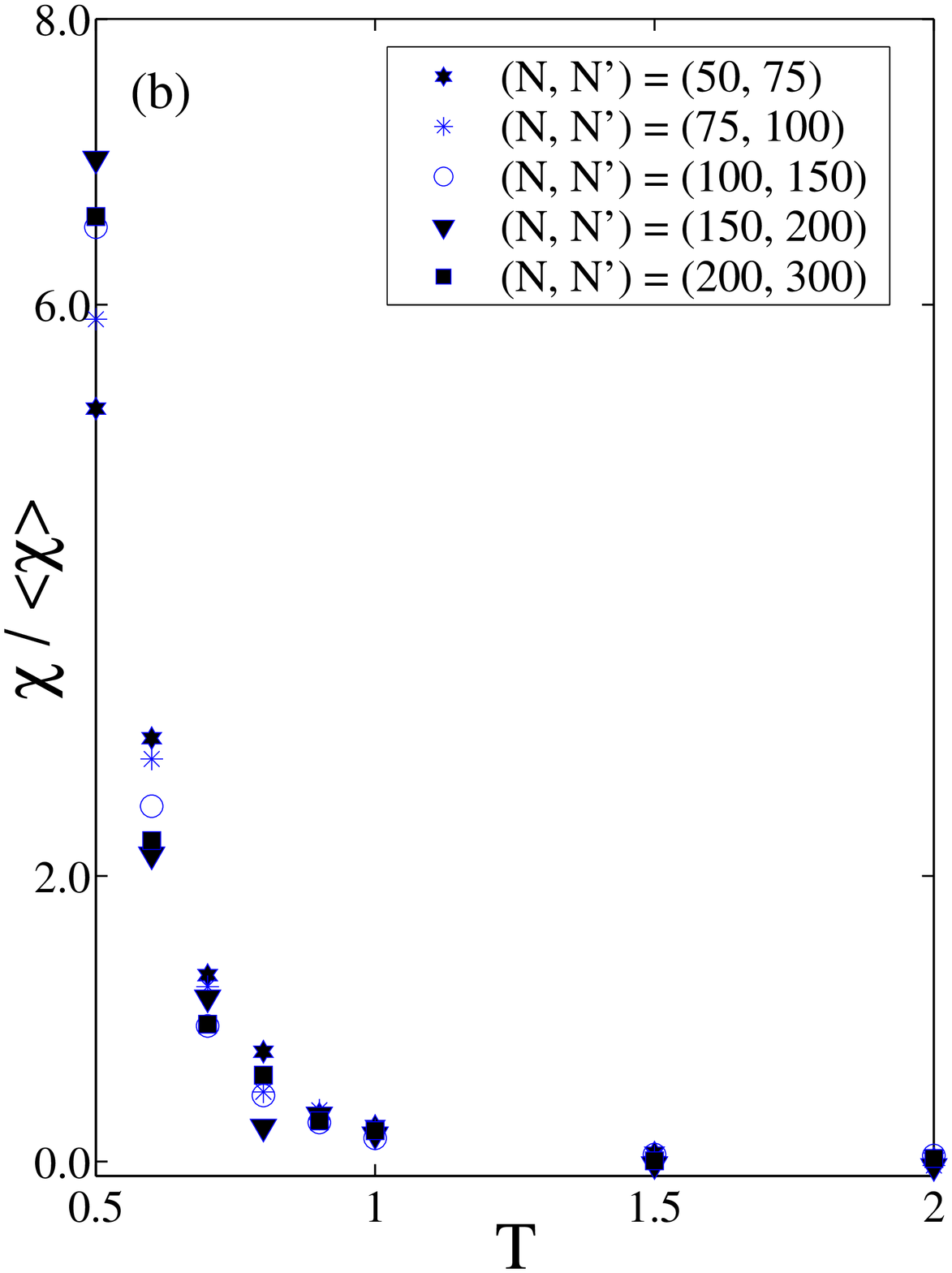}\par
\caption{(a) Average inherent structure energy vs. temperature for the
ROM model. $T_{o} = 0.72$. (b) $ \chi_{N,N^{'}}(T)/\langle
\chi_{N,N^{'}}(T) \rangle $ versus $ T $, where the angle brackets
represent an average over all T.  Curve fitting similar to that of
previous figures yields an approximate onset temperature of
non-mean-field effects of $ T = 0.735 $.}
%(c) $C_{N}(t)$ versus $t$ in
%the $ \alpha $-relaxation range for several different values of $N$ at
%$ T = 0.6.$}
\end{figure}

To determine which portion of the dynamics is sensitive to finite $N$
effects, we have calculated $C_{N}(t)$ for several different values of
$N$ for one particular temperature in the range $T_{o} > T > T_{c}$.
The intermediate time (``$\beta$-relaxation'') regime is not strongly
$N$-sensitive, while the long-time (``$\alpha$-relaxation'') regime
shows strong finite $N$ effects.  This observation is consistent with
a recent study of the dynamics of the finite-sized Random Energy Model
by Ben Arous {\em et al} \cite{BenArous}.  These authors have
demonstrated that dynamical effects beyond the mean-field limit occur
at long times, and that the longest time behavior is compatible with
the predictions of the phenomenological trap model \cite{Monthus}.
Recent computer simulation work has also shown that the long-time
dynamics of atomistic liquids is consistent with an activated-like
trap model precisely in the range $T_{o} > T \geq T_{c}$.  Note that
in the ROM model, for finite $ N, N^{'} $, $ \chi_{N,N^{'}}(T) $
essentially varies continuously through $ T_c $ but changes sharply
near $ T_o $.  Meta-basin dynamics in atomistic simulations also show
a sharp onset of trap-like behavior near $ T_o $, and continuous
variation near $ T_c $ \cite{Doliwa1, Doliwa2, Denny}.  It should also
be noted that recent work by Berthier and Garrahan on kinetic
facilitated models also demonstrates that activated processes set in
at $T_{o}$, where local dynamical heterogeneities begin to
occur\cite{Juanpe2}.

In this paper, we have studied a number of model systems and
demonstrated deep connections between mean-field theory, mode-coupling
theory, and the landscape paradigm. The picture that emerges from this
work and these previous studies is that $ T_o $ marks the edge
boundary for significant barriers that result from a non-mean-field
smearing of the dynamical transition that occurs at $ T_c $ in
mean-field systems. Barriers that would be infinite at $T_{c}$ for an
infinite dimensional liquid become finite, and influence dynamics {\em
not just in the vicinity of $T_{c}$ but at temperatures up to
$T_{o}$}.  Transport involves activation over these barriers, and is
trap-like at very long times.

What, then, can one expect from the idealized version of MCT?  The
analog between $ 1/N $ corrections to mean-field behavior in truncated
mean-field spin glasses and corrections to the idealized MCT suggest
that MCT should {\em always} yield quantitative results for the
$\beta$-relaxation regime.  The remarkable scaling predictions of MCT
in this regime should be unaffected by the mean-field nature of the
approximations inherent in the MCT approach\cite{Gotze}.  While MCT is
able to account for properties such as time-temperature superposition
in the long-time $\alpha$-relaxation regime, finite corrections to MCT
should be noticeable here.  These corrections are connected to
activated (trap-like) behavior that occurs in the range $T_{o} > T $.
Note that the Gaussian trap model also displays time-temperature
superposition in the $\alpha$-relaxation regime.  Clear signs of
dynamic heterogeneities occur in simulations for temperatures in the
range $T_{o} > T \geq T_{c}$\cite{KobGlotz}.  This {\em localized}
heterogeneous motion is likely connected to activated, trap-like
behavior.

While we have shown in this work that there are interesting physical
connections between the temperature $T_{o}$ and the breakdown of
mean-field theory that have not been previously discussed, an open
question still remains regarding the nature of $T_{c}$ as a {\em sharp}
dynamical crossover temperature.  It is instructive to note that in
the p-spin model, $T_{c}$ and $T_{o}$ lie very near each other.
Non-mean-field effects tend to push these temperatures apart.  A
scaled parameter $\tau=\frac{T_{0}-T_{c}}{T_{c}}$ may provide a
measure of how strong the corrections to mean-field behavior are.  In
the binary soft-sphere mixture of Barrat {\em et al.} at unit density
$\tau \approx 0.25$ while for the Lennard-Jones mixture of Kob and
Andersen studied at unit density $\tau \approx 1$.  Interestingly,
dynamical effects such as the appearance of prominent ``hopping''
peaks in the van Hove correlation function $G(r,t)$ \cite{Yama} and
short-time secondary maxima in the non-Gaussian parameter
$\alpha_{2}(t)$ and the nonlinear succeptability $\chi_{4}(t)$ occur
near $T_{c}$ for the soft-sphere mixture but not for the Lennard-Jones
system\cite{Me}.  Perhaps remnants of the behavior expected at $T_{c}$
are only noticeable in systems with small $\tau$ values that are in
some sense closer to idealized mean-field systems\cite{Tomas2}.
Furthermore, it would be interesting to determine if the value of
$\tau$ decreases as the physical dimension of the system increases.  A
systematic study of such open questions is underway.
%It is interesting
%to note, however, that strictly mean-field models may display
%nonlinear succeptabilities that are grossly compatible with the
%heterogeneous motion observed in atomistic simulations.  Indeed, such
%mean-field models may display heterogeneous relaxation.  The
%distinction between the type of heterogeneous relaxation exhibited in
%mean-field spin glasses and real glasses is that heterogeneities in
%liquids are {\em spatially correlated}.  To address this distinction,
%it will be interesting in the future to study the finite $N$ behavior
%of multi-point correlation functions in the ROM, and to see if
%atomistic simulations of particle motion carried out in dimensions
%higher than three show a tendency towards delocalized heterogeneities.
%Currently, we are pursuing both of these ideas.

We gratefully acknowledge useful discussions with G. Biroli,
J.-P. Bouchaud, J.P. Garrahan, P.L. Geissler, A. Heuer and
S. Sastry. We would like to acknowledge NSF support through
grant \#CHE-0134969 and an NSF Fellowship to Y.B.. D.R.R. is an
Alfred P. Sloan Foundation Fellow and a Camille Dreyfus
Teacher-Scholar.

\bibliography{root.bib}

\end{document}